\documentclass[fleqn,usenatbib]{mnras}

\usepackage[T1]{fontenc}

\DeclareRobustCommand{\VAN}[3]{#2}
\let\VANthebibliography\thebibliography
\def\thebibliography{\DeclareRobustCommand{\VAN}[3]{##3}\VANthebibliography}

\usepackage{graphicx}	
\usepackage{amsmath}	
\usepackage{amssymb}	

\title[Prominence detection of RS~CVn]{Prominence detection and chromosphere feature on the prototype RS~CVn of active binary systems}

\author[D. Cao et al.]
             {Dongtao Cao,$^{1, 2}$\thanks{E-mail: dtcao@ynao.ac.cn, shenghonggu@ynao.ac.cn}
             Shenghong Gu,$^{1, 2, 3}$\footnotemark[1]
             U. Wolter,$^{4}$
             M. Mittag,$^{4}$
             J. H. M. M. Schmitt,$^{4}$
\newauthor 
             Dongyang Gao$^{5}$
             and Shaoming Hu$^{5}$
\\
$^{1}$Yunnan Observatories, Chinese Academy of Sciences, Kunming 650216, China\\
$^{2}$Key Laboratory for the Structure and Evolution of Celestial Objects, Chinese Academy of Sciences, Kunming 650216, China\\
$^{3}$School of Astronomy and Space Science, University of Chinese Academy of Sciences, Beijing 101408, China\\
$^{4}$Hamburger Sternwarte, Universit\"{a}t Hamburg, Hamburg D-21029, Germany\\
$^{5}$ Shandong Provincial key Laboratory of Optical Astronomy and Solar-Terrestrial Environment, Institute of Space Sciences, Shandong\\University, Weihai 264209, China}

\date{Accepted XXX. Received YYY; in original form ZZZ}

\pubyear{2015}

\begin{document}
\label{firstpage}
\pagerange{\pageref{firstpage}--\pageref{lastpage}}
\maketitle

\begin{abstract}
We present a study of high-resolution spectra of RS Canum Venaticorum (RS~CVn), a prototype of active binary systems. Our data were obtained from 1998 to 2017 using different telescopes. We analyze the chromospheric activity indicators $\mbox{Ca~{\sc ii}}$ IRT, H$_{\alpha}$, $\mbox{Na~{\sc i}}$ D$_{1}$, D$_{2}$ doublet, $\mbox{He~{\sc i}}$ D$_{3}$, and H$_{\beta}$ using a spectral subtraction technique. The chromospheric emission stems mainly from the K2~IV primary star, while the F5~V secondary star only shows weak emission features in a few of our spectra. We find excess absorption features in the subtracted H$_{\alpha}$ lines and other activity indicators from spectra taken near primary eclipse, which we ascribe to prominence-like material associated with the primary star. We estimate size limits of these tentative prominences based on the geometry of the binary system, and investigate the physical properties of the strongest prominence. An optical flare, characterized by $\mbox{He~{\sc i}}$ D$_{3}$ line emission, together with stronger emission in other activity lines, was detected. The flare energy is roughly comparable to strong flares observed on other RS CVn-type stars. The chromospherically active longitudes of RS~CVn most frequently appear near the two quadratures of the system and display changes between observing runs, which indicates an ongoing evolution of its active regions.
\end{abstract}
\begin{keywords}
	         stars: activity --
                   stars: chromospheres --
                   stars: flare --
                   stars: circumstellar matter --
                   stars: binaries: spectroscopic
\end{keywords}
\section{Introduction}
\indent
A definition of RS CVn-type binary systems was first performed by \citet{Hall1976} and subsequently refined by \citet{Fekel1986}: a binary system which has at least one cool component showing strong magnetic activity in several forms. This activity manifests itself as strong photometric variability caused by large starspots, chromospheric emission, transition region emission, and coronal radiation. It is commonly accepted that all of these active phenomena arise from a powerful magnetic dynamo generated by the interplay between turbulent motions in the convection zone and the stellar differential rotation. At Yunnan Observatories, we are performing a long-term high-resolution spectroscopic monitoring project that covers several RS CVn-type systems to study their magnetic activity phenomena \citep[e.g.][]{Gu2002, Cao2015, Cao2019}. In the present paper, we focus on the prototype system RS~CVn itself.\\
\indent
RS~CVn (= HD~114519 = BD+36$\degr$2344) is a double-lined, totally eclipsing binary system consisting of a K2~IV primary and a F5~V secondary \citep{Reglero1990,Rodono2001}. The system has a period of about 4.798~days in an almost circular orbit with an inclination of $i$~=~85$\degr$ \citep{Catalano1974,Eaton1993}. Table~\ref{tab1} summarizes RS~CVn's properties, compiled from \citet{Eaton1993} and \citet{Rodono2001}.\\
\indent
RS~CVn shows a significant distortion wave in its photometric light curve, caused by starspots on the surface of the K2~IV primary star \citep[e.g.][]{Eaton1993, Heckert1995, Rodono1995, Rodono2001}. Based on long-term light curves obtained between 1949 and 1993, \citet{Rodono1995} analyzed its starspot evolution, possible activity cycles and orbital period variations. \citet{Rodono2001} further determined accurate photometric parameters using a long-term sequence of light curves of RS CVn, taking into account the light curve distortions caused by starspots. More recently, using Doppler imaging, \citet{Xiang2020} reconstructed the starspot distribution on the primary star and estimated its surface differential rotation based on images derived from two consecutive rotational cycles.\\
\indent
Chromospheric activity of RS~CVn has been studied by several authors, e.g. \citet{Popper1988}, \citet{Reglero1990}, \citet{Fernandez1986, Fernandez1994}, and \citet{Montes1996a}. RS~CVn shows pronounced chromospheric emission in $\mbox{Ca~{\sc ii}}$ H \& K and other activity lines, the emission features are mainly associated with the K2~IV primary star \citep{Reglero1990, Montes1996a}.\\
\indent
So far, we know only little about the magnetic activity of RS~CVn's outer atmosphere, in particular in relation to its chromosphere. Here we present the detection of prominence-like structures on RS~CVn and the results of our chromospheric activity study based on the $\mbox{Ca~{\sc ii}}$ IRT, H$_{\alpha}$, $\mbox{Na~{\sc i}}$ D$_{1}$, D$_{2}$ doublet, $\mbox{He~{\sc i}}$ D$_{3}$, and H$_{\beta}$ lines. Our study makes use of a large set of high-resolution spectra obtained during several observing runs between 1998 and 2017. This includes data from a joint observation campaign using telescopes located in China and Mexico in 2017~April, designed to get a dense short-term phase coverage. In Section~2, we provide details of our observations and data reduction. Our analysis of chromospheric activity indicators is described in Section~3. In Section~4, we describe in detail the different activity phenomena of RS~CVn during our observations, including prominence-like events, optical flares, and chromospheric activity variations. Finally, we state our conclusions in Section~5.\\
\begin{table}
\centering
\caption{RS~CVn system parameters.}
\tabcolsep 0.10cm
\label{tab1}
\begin{tabular}{lcc}
\hline
\multicolumn{1}{l}{Parameter} &
\multicolumn{1}{c}{Primary} &
\multicolumn{1}{c}{Secondary}\\
\hline
Spectral type&K2~IV&F5~V\\
M/M$_{\sun}$&1.44&1.39\\
R/R$_{\sun}$&3.85&1.89\\
Semi-major axis&\multicolumn{2}{c}{16.9R$_{\sun}$}\\
Orbital inclination&\multicolumn{2}{c}{85$\degr$}\\
Orbital period&\multicolumn{2}{c}{4$^{d}.797695$}\\
Mid-primary eclipse (HJD)$^a$&\multicolumn{2}{c}{2, 448, 379.1993}\\
\hline
\multicolumn{3}{c}{$^a$ Primary eclipse is defined as phase zero, where the secondary}\\
\multicolumn{3}{c}{component is totally eclipsed by the primary one.}\\
\end{tabular}
\end{table}
\section{Observations and data reduction}
\begin{figure*}
\centering
\includegraphics[width=17.5cm, height=17.5cm,angle=270]{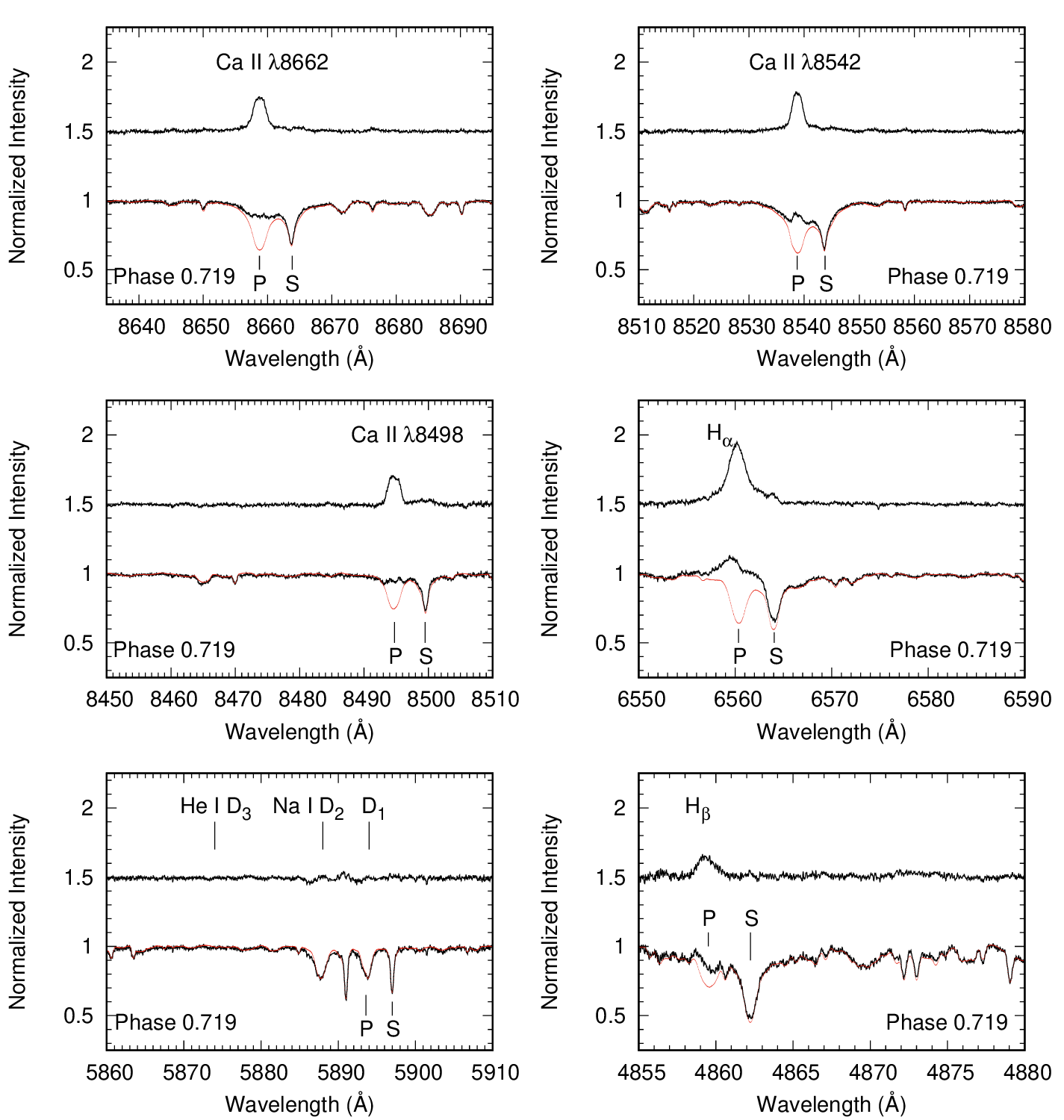}
\caption{Example of the observed, synthesized, and subtracted spectra for the annotated spectral line regions, shown for one spectrum obtained at at Xinglong (2017 Nov~30). In each panel, the lower solid-line indicates the observed spectrum, the dotted red line represents the synthesized spectrum constructed from spectra of two template stars, and the upper graph shows the resulting subtraction spectrum, shifted for better visibility. "P" and "S" mark the positions of the chromospheric activity lines for the primary and secondary components of RS~CVn, respectively.}
\label{Fig1}
\end{figure*}

\begin{table}
\centering
\caption{RS~CVn observing log.}
\tabcolsep 0.3cm
\label{tab2}
\begin{tabular}{lccccr}
\hline
\multicolumn{1}{l}{Date} &
\multicolumn{1}{c}{HJD} &
\multicolumn{1}{c}{Phase}&
\multicolumn{1}{r}{Exp.time}\\
   & (2,450,000+)&&(s)\\
\hline
\multicolumn{4}{c}{1998 Mar, Xinglong~2.16~m}\\
1998 Mar 14&0887.3212&0.776&3600\\
1998 Mar 14&0887.3635&0.785&3600\\
&&&\\
\multicolumn{4}{c}{2000 Feb, Xinglong~2.16~m}\\
2000 Feb 20&1595.3134&0.346&1800\\
2000 Feb 20&1595.3352&0.350&1800\\
&&&\\
\multicolumn{4}{c}{2004 Feb, Xinglong~2.16~m}\\
2004 Feb 03&3039.3954&0.341&3600\\
2004 Feb 05&3041.2822&0.734&2760\\
2004 Feb 05&3041.3664&0.752&2400\\
2004 Feb 07&3043.3250&0.160&3600\\
2004 Feb 08&3044.3839&0.380&3600\\
2004 Feb 09&3045.3200&0.576&3600\\
&&&\\
 \multicolumn{4}{c}{2016 Jan, Xinglong~2.16~m}\\
2016 Jan 22&7410.3077&0.385&3600\\
2016 Jan 23&7411.3284&0.598&3600\\
2016 Jan 23&7411.3775&0.608&3600\\
2016 Jan 24&7412.2358&0.787&1800\\
2016 Jan 24&7412.2889&0.798&1800\\
2016 Jan 24&7412.3753&0.816&1800\\
2016 Jan 24&7412.3985&0.821&1800\\
2016 Jan 25&7413.2727&0.003&1800\\
2016 Jan 25&7413.3265&0.014&1800\\
2016 Jan 26&7414.2849&0.214&1800\\
2016 Jan 26&7414.3373&0.225&1800\\
2016 Jan 26&7414.3904&0.236&1800\\
2016 Jan 26&7414.4136&0.241&1800\\
2016 Jan 28&7416.2773&0.629&1800\\
2016 Jan 28&7416.3006&0.634&1800\\
2016 Jan 29&7417.2814&0.838&1800\\
2016 Jan 29&7417.3048&0.843&1800\\
2016 Jan 30&7418.3252&0.056&1800\\
2016 Jan 30&7418.3484&0.061&1800\\
2016 Jan 30&7418.3718&0.066&1800\\
2016 Jan 30&7418.3950&0.071&1800\\
2016 Jan 31&7419.2627&0.251&1800\\
2016 Jan 31&7419.2859&0.256&1800\\
2016 Jan 31&7419.3576&0.271&1800\\
2016 Jan 31&7419.3809&0.276&1800\\
&&&\\
\multicolumn{4}{c}{2017 Apr, Weihai~1~m}\\
2017 Apr 17&7861.0953&0.344&3600\\
2017 Apr 18&7862.0654&0.546&3600\\
2017 Apr 18&7862.1074&0.555&3600\\
2017 Apr 18&7862.1500&0.564&3600\\
2017 Apr 18&7862.2457&0.584&3600\\
2017 Apr 18&7862.2876&0.593&3600\\
2017 Apr 18&7862.3297&0.601&3600\\
&&&\\
\multicolumn{4}{c}{2017 Apr, TIGRE~1.2~m}\\
2017 Apr 14&7857.6464&0.625&800\\
2017 Apr 14&7857.6914&0.635&800\\
2017 Apr 14&7857.7343&0.644&800\\
2017 Apr 14&7857.7748&0.652&800\\
2017 Apr 14&7857.8177&0.661&800\\
2017 Apr 14&7857.8589&0.670&800\\
2017 Apr 14&7857.9406&0.687&800\\
2017 Apr 15&7858.6400&0.832&800\\
2017 Apr 16&7859.7464&0.063&800\\
2017 Apr 16&7859.7905&0.072&800\\
\end{tabular}
\end{table}
\begin{table}
\centering
\contcaption{}
\tabcolsep 0.3cm
\label{tab2}
\begin{tabular}{lccccr}
\hline
   \multicolumn{1}{l}{Date}&
   \multicolumn{1}{c}{HJD}&
   \multicolumn{1}{c}{Phase}&
   \multicolumn{1}{r}{Exp.time}\\
   & (2,450,000+)&&(S)\\
\hline
2017 Apr 16&7859.8354&0.081&800\\
2017 Apr 16&7859.8808&0.091&800\\
2017 Apr 16&7859.9340&0.102&800\\
2017 Apr 17&7860.6080&0.243&800\\
2017 Apr 17&7860.6722&0.256&800\\
2017 Apr 17&7860.7390&0.270&800\\
2017 Apr 17&7860.8788&0.299&800\\
2017 Apr 18&7861.6112&0.452&800\\
2017 Apr 19&7862.7679&0.693&800\\
2017 Apr 19&7862.8081&0.701&800\\
2017 Apr 19&7862.8686&0.714&800\\
2017 Apr 19&7862.9071&0.722&800\\
2017 Apr 20&7863.5836&0.863&800\\
2017 Apr 20&7863.6263&0.872&800\\
2017 Apr 20&7863.6656&0.880&800\\
2017 Apr 20&7863.7049&0.888&800\\
2017 Apr 20&7863.7447&0.896&800\\
2017 Apr 20&7863.7830&0.904&800\\
2017 Apr 20&7863.8261&0.913&800\\
2017 Apr 20&7863.8789&0.924&800\\
2017 Apr 20&7863.9173&0.932&800\\
2017 Apr 21&7864.6062&0.076&800\\
2017 Apr 21&7864.6449&0.084&800\\
2017 Apr 21&7864.6829&0.092&800\\
2017 Apr 21&7864.7222&0.100&800\\
2017 Apr 21&7864.7610&0.108&800\\
2017 Apr 21&7864.7995&0.116&800\\
2017 Apr 21&7864.8452&0.126&800\\
2017 Apr 21&7864.8839&0.134&800\\
2017 Apr 21&7864.9236&0.142&800\\
&&&\\
\multicolumn{4}{c}{2017 Nov--Dec, Xinglong~2.16~m}\\
2017 Nov 28&8086.3636&0.298&1800\\
2017 Nov 28&8086.3868&0.302&1800\\
2017 Nov 28&8086.4100&0.307&1800\\
2017 Nov 30&8088.3846&0.719&1800\\
2017 Nov 30&8088.4078&0.724&1800\\
2017 Dec 01&8089.3822&0.927&1800\\
2017 Dec 01&8089.4054&0.932&1800\\
2017 Dec 07&8095.3590&0.172&1800\\ 
2017 Dec 07&8095.3822&0.177&1800\\  
2017 Dec 07&8095.4054&0.182&1800\\ 
2017 Dec 08&8096.3461&0.378&1800\\  
2017 Dec 08&8096.3692&0.383&1800\\ 
2017 Dec 08&8096.3924&0.388&1800\\ 
2017 Dec 08&8096.4156&0.393&1800\\  
2017 Dec 09&8097.3639&0.590&1800\\ 
2017 Dec 09&8097.3870&0.595&1800\\ 
2017 Dec 09&8097.4104&0.600&1800\\  
2017 Dec 10&8098.3282&0.791&1800\\  
2017 Dec 10&8098.3514&0.796&1800\\  
2017 Dec 10&8098.3746&0.801&1800\\  
2017 Dec 10&8098.3978&0.806&1800\\ 
2017 Dec 10&8098.4211&0.811&1800\\ 
2017 Dec 11&8099.3624&0.007&1800\\ 
2017 Dec 11&8099.3856&0.012&1800\\  
2017 Dec 11&8099.4088&0.017&1800\\  
\hline
\end{tabular}
\end{table}
\indent
During the runs of Mar.~1998, Feb.~2000, and Feb.~2004, we used the coud\'{e} echelle spectrograph (CES, \citealt{Zhao2001}) mounted on the 2.16-m telescope at the Xinglong station, National Astronomical Observatories, Chinese Academy of Sciences. The spectrograph covered the wavelength range of 5600--9000~\AA~with an average resolving power of R~=~$\lambda$/$\Delta\lambda$~$\simeq$~37000, and the data were recorded on a $1024 \times 1024$ pixel CCD. The fiber-fed high-resolution spectrograph HRS was later installed on the 2.16-m telescope, which we used in Jan.~2016 and Nov--Dec.~2017. HRS produces spectra with a resolving power of R~$\simeq$~48000 on a wavelength range of 3900--10000~\AA, using a $4096 \times 4096$ pixel CCD.\\
\indent
During 2017, Apr~14 to~21, furthermore, we made a joint observation by using the 1-m telescope at the Weihai Observatory of Shandong University, China \citep{Gao2016} and the TIGRE 1.2-m telescope of Hamburg Observatory, Germany \citep{Schmitt2014}. The Weihai Echelle Spectrograph (WES) is fiber-fed and covers the wavelength range 3800--9000~\AA~with an average resolving power of R~$\simeq$~50000. TIGRE is a robotic telescope located near Guanajuato in Central Mexico and equipped with the fiber-fed spectrograph HEROS. Its spectra have a resolving power of R~$\simeq$~20000, covering a wavelength range of 3800--5700~\AA~and 5800--8800~\AA~in its blue and red arm, respectively.\\
\indent
Table~\ref{tab2} lists our observations of RS~CVn, which including the used instrument, the observing date, HJD, orbital phase, and exposure time. The orbital phases were calculated using the ephemerides given in Table~\ref{tab1}. Besides our target RS~CVn, during each run, we observed a few rapidly rotating early-type stars, used as telluric absorption line templates. Finally, we observed several slowly rotating, inactive stars with the same spectral type and luminosity class as the components of RS~CVn. They are required for the spectral subtraction technique described in Section~3.\\
\indent
TIGRE observations are automatically reduced by the TIGRE pipeline \citep{Mittag2010}. We normalized the extracted and wavelength calibrated spectra by using a low-order polynomial fit to the observed continuum with the IRAF$^{1}$\footnotetext[1]{IRAF is distributed by the National Optical Astronomy Observatories, which is operated by the Association of Universities for Research in Astronomy (AURA), Inc., under cooperative agreement with the National Science Foundation.} package.\\
\indent
All other observations were completely reduced with the IRAF package, following the standard procedures and using Th-Ar spectra of the corresponding nights for wavelength calibration. As a final step, also these spectra were normalized using low-order polynomial fit to the observed continuum.\\ 
\indent
Some of our observations during Nov--Dec.~2017, obtained at Xinglong, and Apr.~2017 taken by TIGRE, were heavy contaminated by telluric absorption lines in the chromospheric activity regions of interest. We eliminated them using the spectra of two rapidly rotating early-type stars HR~989 (B5~V, $vsini$ = 298~km~s$^{-1}$) and HR~7894 (B5~IV, $vsini$ = 330~km~s$^{-1}$), respectively, with an interactive procedure in the IRAF package.\\
\indent
Examples of the $\mbox{Ca~{\sc ii}}$ IRT, H$_{\alpha}$, $\mbox{Na~{\sc i}}$ D$_{1}$, D$_{2}$ doublet, $\mbox{He~{\sc i}}$ D$_{3}$, and H$_{\beta}$ line profiles of RS~CVn are displayed in Figure~\ref{Fig1}.\\
\section{Analysis of chromospheric activity indicators}
\indent
We simultaneously analyze different chromospheric activity indicators, namely $\mbox{Ca~{\sc ii}}$ IRT, H$_{\alpha}$, $\mbox{Na~{\sc i}}$ D$_{1}$, D$_{2}$ doublet, $\mbox{He~{\sc i}}$ D$_{3}$, and H$_{\beta}$ lines, formed in a wide range of atmospheric heights. To separate the contribution from the photosphere absorption profile in these lines, we apply a spectral subtraction technique, making use of the STARMOD program \citep{Barden1985, Montes1997, Montes2000}. STARMOD synthesizes a spectrum, later subtracted, by rotationally broadening, RV-shifting, and adaptively weighting spectra of two suitable template stars. These template stars are chosen as inactive, but having the same spectral-type and luminosity class as the components of the system. Thus, the synthesized spectrum approximates the non-active state of the binary, and the subtraction between the observed and synthesized spectra produces the activity contribution as excess emission and/or absorption, relative to the presumed inactive state.\\
\indent 
For the observations of Jan. 2016 and Nov--Dec. 2017, taken at Xinglong, we use spectra of HR~8088 (K2~IV) and HR~3262 (F6~V) as templates for the primary and secondary star, respectively. The $vsini$ of each binary component is determined by STARMOD using these template spectra, resulting in an average $vsini$ of 45 km~s$^{-1}$ for the primary star and 14 km~s$^{-1}$ for the secondary one, based on a high signal-to-noise ratio (SNR) spectral window spanning many photospheric absorption lines, observed at phases where the two components of the system are well separated. The obtained $vsini$ values are in good agreement with the results of 42 $\pm$ 3~km~s$^{-1}$ \& 11 $\pm$ 2~km~s$^{-1}$ estimated by \citet{Strassmeier1990}, and 44.9 $\pm$ 1~km~s$^{-1}$ \& 12.4 $\pm$ 0.5~km~s$^{-1}$ by \citet{Xiang2020}. The adopted intensity ratios (primary/secondary) used by STARMOD are 0.57/0.43 for the $\mbox{Ca~{\sc ii}}$ $\lambda$8662 spectral region, 0.56/0.44 for the $\mbox{Ca~{\sc ii}}$ $\lambda$8542 spectral region, 0.55/0.45 for the $\mbox{Ca~{\sc ii}}$ $\lambda$8498 spectral region, 0.5/0.5 for the H$_{\alpha}$ spectral region, 0.47/0.53 for the $\mbox{Na~{\sc i}}$ D$_{1}$ and D$_{2}$ doublet, $\mbox{He~{\sc i}}$ D$_{3}$ spectral region, finally 0.43/0.57 for the H$_{\beta}$ spectral region.\\
\indent
The same template stars, HR~8088 and HR~3262, are also used for the observations of Apr.~2017, obtained at Weihai, as well as the observations of Mar.~1998, Feb.~2000, and Feb.~2004 obtained at Xinglong. We use the same $vsini$ values and intensity weight ratios as given above, with the exception of the H$_{\beta}$ line region for the Weihai observations, because of very low SNR, and the Xinglong observations, where the line is not covered. Also, during the 1998 observing run, the $\mbox{Ca~{\sc ii}}$ $\lambda$8498 line was not covered by the spectra due to an echelle setup change.\\
\indent
Because the template stars used above were not observed by the TIGRE telescope, we use spectra of two other inactive stars HR~5227 (K2~IV) and HR~8697 (F6~V) as templates instead. The $vsini$ values of 45 \& 14 km~s$^{-1}$ are used, and the resulting and adopted intensity weight ratios are 0.59/0.41 for the $\mbox{Ca~{\sc ii}}$ IRT spectral region, 0.51/0.49 for the H$_{\alpha}$ spectral region, and 0.44/0.56 for the H$_{\beta}$ spectral region. Also for some of the TIGRE observations, H$_{\beta}$ could not be analyzed because of too low SNRs.\\
\indent
During eclipses of RS~CVn's components, the intensity weights of the two components change with orbital phase, and the line profiles are distorted -- a situation that STARMOD can not model. Since some of our observations were obtained during eclipse, the resulting spectra have been excluded from our analysis, they are not listed in Table~\ref{tab1}. This applies with the exception of spectra observed during total primary eclipse, i.e. when the secondary is completely occulted by the primary, such as observations on 2017 Dec~11. In these cases, we do use the template spectrum of the K2~IV alone to perform our analysis.\\
\indent
As a typical example, we show the chromospherically sensitive lines of one spectrum and illustrate the above described processing in Fig.~\ref{Fig1}. As expected, RS~CVn shows pronounced chromospheric emission in all analyzed lines, mainly associated with the primary star. Moreover, for several observations of 2017 November--December and some observations of other observing runs, there are obvious excess emission features associated with the secondary star of RS~CVn in the subtracted spectra (especially in the H$_{\alpha}$ line), which indicates that the secondary star is also active in the system, although much weaker.\\
\indent
The equivalent widths (EWs) of the subtracted $\mbox{Ca~{\sc ii}}$ IRT, H$_{\alpha}$, and H$_{\beta}$ line profiles are measured with the SPLOT task in the IRAF package, as described in our previous papers \citep{Cao2015, Cao2017}, they are summarized in Table~\ref{tab3} together with their estimated uncertainties, where we also provide the ratios of the EW($\lambda$8542)/EW($\lambda$8498) and the E$_{H{\alpha}}$/E$_{H{\beta}}$ for some of our observations. The E$_{H{\alpha}}$/E$_{H{\beta}}$ ratios are calculated from the EW(H$_{\alpha}$)/EW(H$_{\beta}$) values with the correction:
\begin{equation}
\frac{E_{H{\alpha}}}{E_{H{\beta}}} = \frac{EW(H_{\alpha})}{EW(H_{\beta})}*0.2444*2.512^{(B-R)}
\end{equation}
given by \citet{Hall1992}, which takes into account the absolute flux density in H$_{\alpha}$ and H$_{\beta}$ lines, and the color index; we use $B-R$~=~0.81 for the calculation here.\\
\indent
The EW$_{8542}$/EW$_{8498}$ ratios thus obtained lie in the range of 1.0--2.5, consistent with the ratios found in solar plages ($\sim$1.5--3; \citealt{Chester1991}) as well as several other chromospherically activity stars \citep[e.g.][]{Montes2000, Gu2002, Cao2015, Cao2020}, thereby suggesting that the $\mbox{Ca~{\sc ii}}$ IRT line emission arises predominantly from plage-like regions.\\
\begin{figure*}
\centering
\includegraphics[width=11cm,height=8.75cm,angle=270]{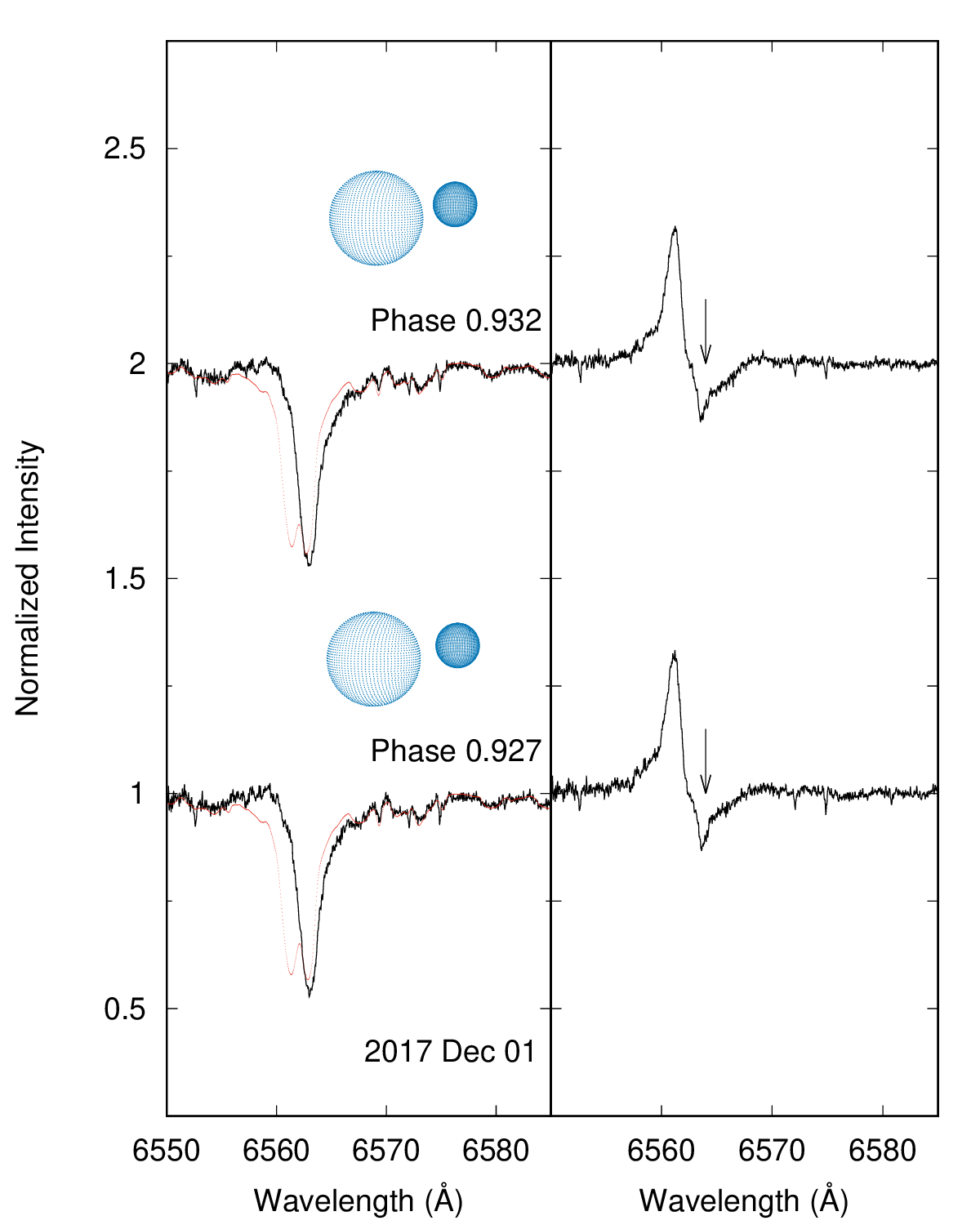}
\includegraphics[width=11cm,height=8.75cm,angle=270]{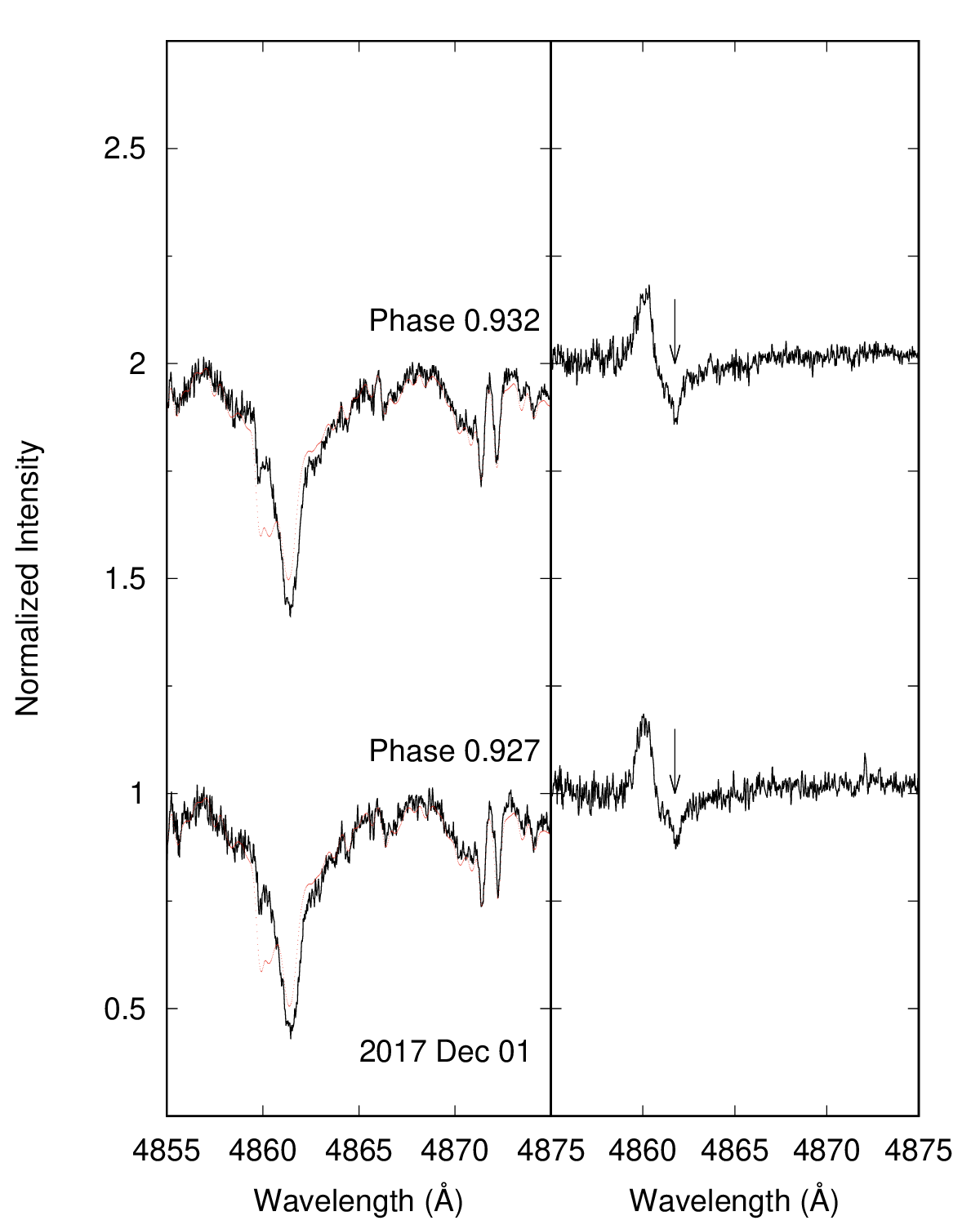}
\caption{Excess absorption features in H$_{\alpha}$ (left panel) and H$_{\beta}$ (right panel), observed at phases 0.927 and 0.932 on 2017 December~1. The left side of each panel shows the observed (solid) and synthesized (dotted) spectra, while the right side contains the subtracted spectra. The geometry of the system at each phase is also shown. Arrows indicate excess absorption features in the subtracted spectra.}
\label{Fig2}
\end{figure*}
\begin{figure}
\centering
\includegraphics[width=8.5cm,height=10.7cm]{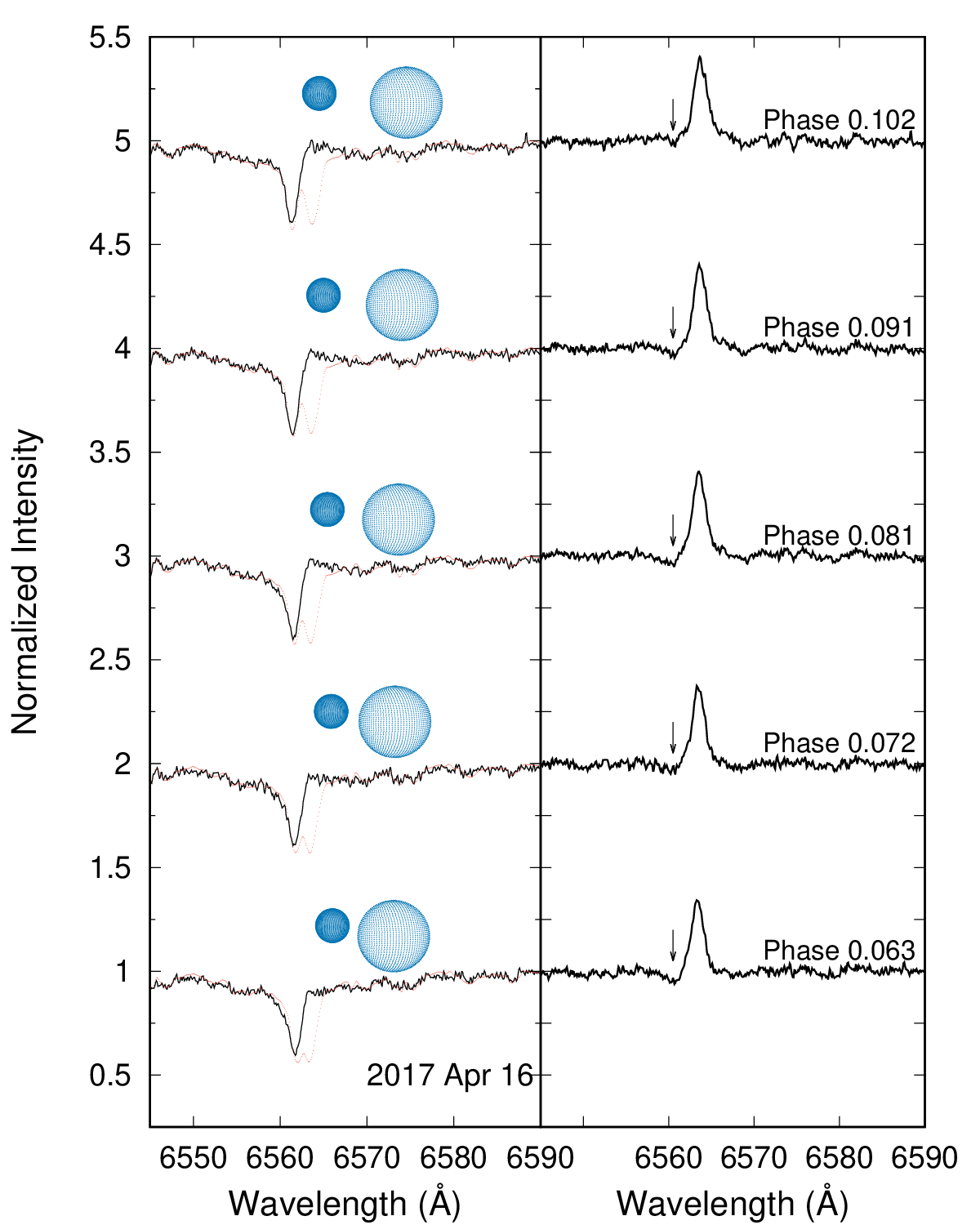}
\includegraphics[width=8.5cm,height=10.7cm]{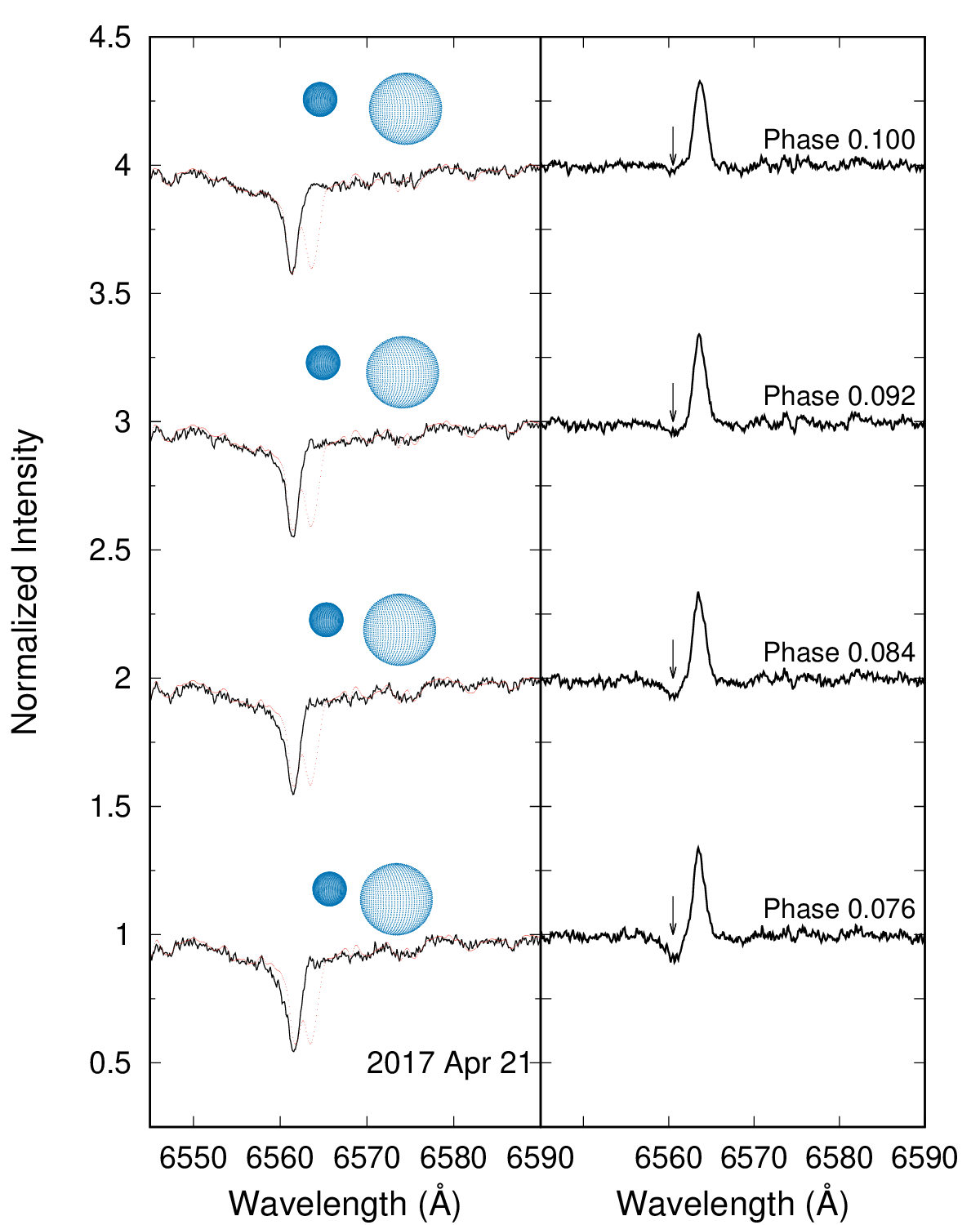}
\caption{Same as Fig.~\ref{Fig2}, for excess absorption features in the H$_{\alpha}$ line profiles observed on 2017 April 16 (top) and 21 (bottom).}
\label{Fig3}
\end{figure}
\section{Results and discussion}
\subsection{Prominence-like absorption features}
\subsubsection{Excess absorption features}
\indent
At phases near primary eclipse, some of our RS~CVn spectra show excess absorption features, visible in the STARMOD-subtracted spectral profiles, most strongly in the Balmer lines.\\
\indent
As shown in Fig.~\ref{Fig2}, for the observations on 2017 December~01 at phases 0.927 and 0.932, shortly before the primary star began to eclipse the secondary one, an excess absorption feature appeared in the red wing of the subtracted H$_{\alpha}$ and H$_{\beta}$ line profiles. Simultaneously, an absorption feature appeared in the $\mbox{He~{\sc i}}$ D$_{3}$ line region, in both the observed and subtracted spectra. We had previously observed similar absorption features in the same lines of the RS CVn-type star SZ~Psc \citep{Cao2019, Cao2020}.\\
\indent
From Fig.~\ref{Fig3}, moreover, it can be seen that there were excess absorption features presented in the blue wing of the subtracted H$_{\alpha}$ line profile at several phases on 2017 April~16, and again on 2017 April~21. These absorption features again occurred at orbital phases just after primary eclipse, when the projected separation between two components gradually becomes larger, repeating after one orbital cycle. In the subtracted spectra, furthermore, we note that the intensity of excess absorption gradually decayed until the features disappeared (indicated by the arrows in Fig.~\ref{Fig3}). Different from what we found on 2017 December~01, however, there were no significant excess absorption features in the blue wings of H$_{\beta}$ line, probably because of much lower SNRs.\\
\indent
We attribute these excess absorption features to prominence-like material associated with the primary star, thereby absorbing radiation from the secondary star near the primary eclipse. This occurs during these phases because the materials lie near the secondary star in velocity space, which, in turn, is located behind the primary star along the line of our sight. Stellar prominences have been detected as transient absorption features passing through the rotationally broadened H$_{\alpha}$ line profiles on a number of rapidly rotating single stars such as AB Dor \citep{Collier1989}, Speedy~Mic \citep{Jeffries1993, Dunstone2006a, Wolter2008}, HK~Aqr \citep{Byrne1996}, RE 1816+514 \citep{Eibe1998}, PZ~Tel \citep{Barnes2000}, and RX J1508.6--4423 \citep{Donati2000}. Transient absorption features are thought to originate from cool clouds of mostly neutral material, magnetically supported above the stellar photosphere and forced to corotate with the star in a manner reminiscent of solar prominences, which then scatter the underlying chromospheric emission out of the line of sight as they transit the stellar disk \citep{Collier1989}. Actually, Balmer excess absorptions, interpreted in terms of prominence-like features, have even been detected in several RS~CVn-type stars, mainly double-lined eclipsing binary systems \citep{Hall1990, Hall1992, Frasca2000, Cao2019, Cao2020}. Furthermore, we detected such features several times in the RS~CVn-type binary star SZ~Psc \citep{Cao2019, Cao2020}, including cases apparently caused by flare-related prominence activation and post-flare loops \citep{Cao2019}.\\
\indent
The $E_{H{\alpha}}/E_{H{\beta}}$ values have repeatedly been used as diagnostics for discriminating between prominence-like and plage-like structures. \citet{Huenemoerder1987} found that low ratios in RS CVn-type stars are caused by plage-like regions, while prominence-like structures have high values. \citet{Buzasi1989} also developed a NLTE radiative transfer model and concluded that the low ratios ($\sim$ 1--2) could be achieved both in plages and prominences viewed against the disk, but high values ($\sim$ 3--15) could only be achieved in prominence-like structures viewed off the stellar limb. For RS~CVn, the ratios we have obtained are in the range of 1--3 for most of our observations (see Table~\ref{tab3}), which suggests that there are plage-like regions in the chromosphere, in agreement with the results of the EW$_{8542}$/EW$_{8498}$. There are much higher $E_{H{\alpha}}/E_{H{\beta}}$ values for the observations at phase 0.927 and 0.932 during 2017 Nov--Dec, which exhibit excess absorption features in the subtractions, this should be because of prominence-like structure.\\
\indent
On the other hand, for the excess absorptions present in the subtracted H$_{\alpha}$ lines near primary eclipse in 2017 April, the $E_{H{\alpha}}/E_{H{\beta}}$ ratios are not high. Furthermore, the excess H$_{\alpha}$ absorptions are not strong here and no obvious excess absorptions appear in H$_{\beta}$, though potentially hidden by low SNRs. This implies that the prominence activity was much weaker during those observations.\\
\subsubsection{Prominence heights}
\indent
Since the geometry of RS~CVn is well established, we can try to infer approximate limits on the physical extent of prominences, using the parameters listed in Table~1. For the excess absorption features observed on 2017 April~16 and 21, both of them appeared and disappeared around a similar phase (see Fig.~\ref{Fig3}). Large stellar prominences may have lifetimes spanning several stellar rotations \citep{Donati1999}, therefore, both can be expected to be caused by the same prominence. At phase 0.102 on 2017 April~16, the projected separation between the limbs of the two stellar components is about 4.32~$R_{\sun}$, which suggests that the prominence extends at least up to this value from the surface of the primary star. Furthermore, the separation between the limbs of the two components is about 0.74~$R_{\sun}$ at phase 0.063, and therefore the separation between the leading limb of the primary star and the far limb of the secondary star is 0.74~$R_{\sun}$~+~2~*~$R_{sec}$~=~4.52~$R_{\sun}$, which can also be regarded as the upper limit on the prominence's height from the surface of the primary, which would still allow it to be projected on the stellar disk of the secondary. Hence, we conclude that this prominence probably extends from the surface of the primary star in the range between 4.32~$R_{\sun}$ ($\simeq$~1.12~$R_{pri}$) and 4.52~$R_{\sun}$ ($\simeq$~1.17~$R_{pri}$).\\
\indent
Following the same line of reasoning, for the excess absorption detected on 2017 December~1, the separation between the limbs of the two system components at phase 0.927 is about 1.74~$R_{\sun}$ in the plane of the sky, which again means that the prominence extends at least up to this height above the surface of the primary. At phase 0.932, the separation between the limbs of the two components is about 1.28~$R_{\sun}$ and the separation between the leading limb of the primary and the far limb of the secondary is 1.28~$R_{\sun}$~+~2~*~$R_{sec}$~=~5.06~$R_{\sun}$. To summarize, this prominence structure probably extends from the surface of the primary star somewhere between 1.74~$R_{\sun}$ ($\simeq$~0.45~$R_{pri}$) and 5.06~$R_{\sun}$ ($\simeq$~1.31~$R_{pri}$).\\
\indent
In single stars, stellar prominences reach heights of several stellar radii, they usually lie near or beyond the corotation radius. For example, the distances of many prominences from the stellar rotation axis are mainly between 3 and 5 stellar radii for the star AB Dor with the corotation radius at 2.7 stellar radii \citep{Collier1989}. For the primary star of RS~CVn, however, its Keplerian corotation radius ($R_{k}~=~\sqrt[3]{GM/{\Omega}^{2}}$) is calculated to be about 3.3~$R_{pri}$ (2.3~$R_{pri}$ from the surface). Therefore, according to our estimate, prominence-like structures of RS~CVn appear to be formed within the corotation radius of the primary star, indicating a different behavior, compared to stellar prominences in single stars.\\
\begin{figure}
\centering
\includegraphics[width=6.25cm,height=8.5cm,angle=270]{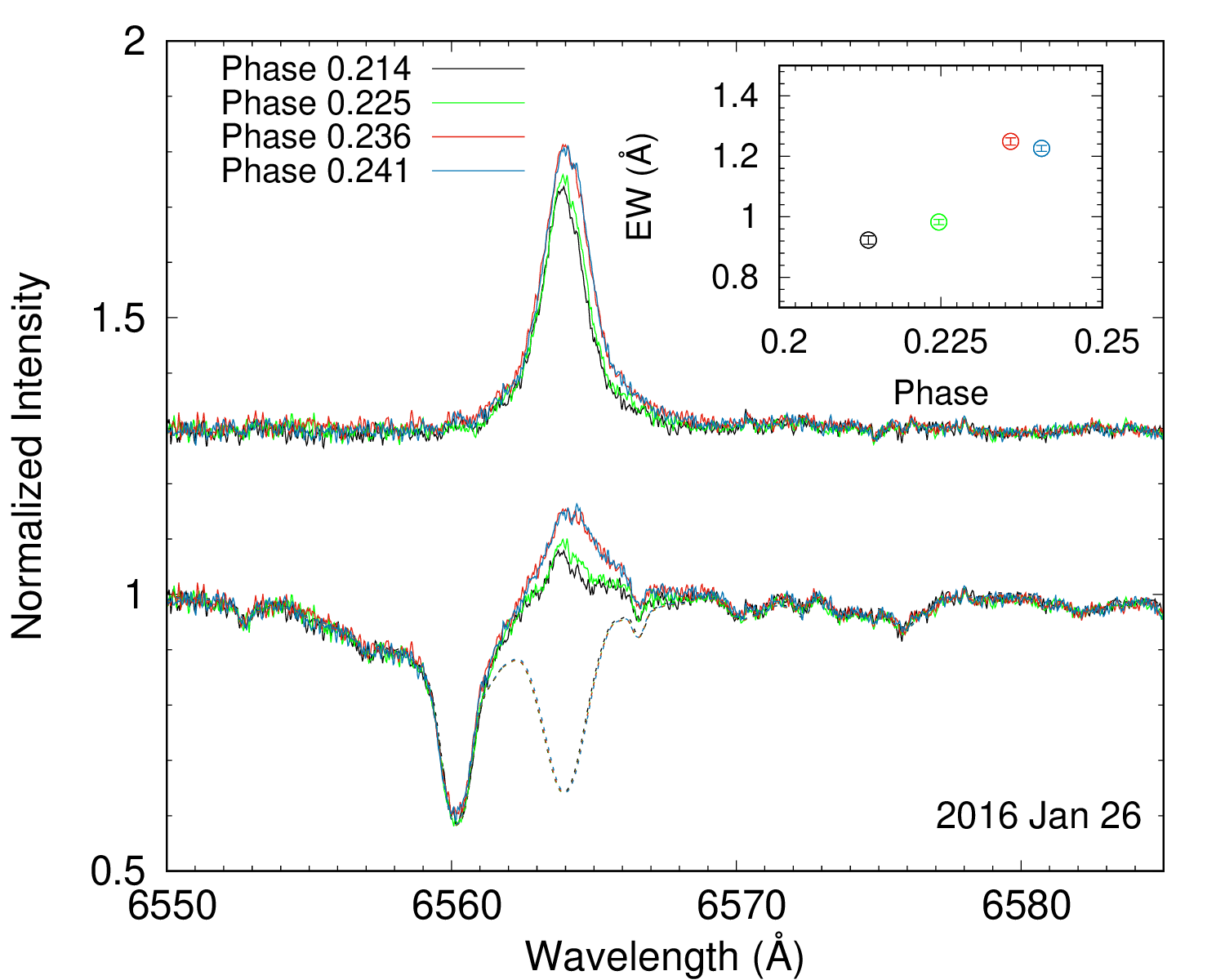}
\includegraphics[width=6.25cm,height=8.5cm,angle=270]{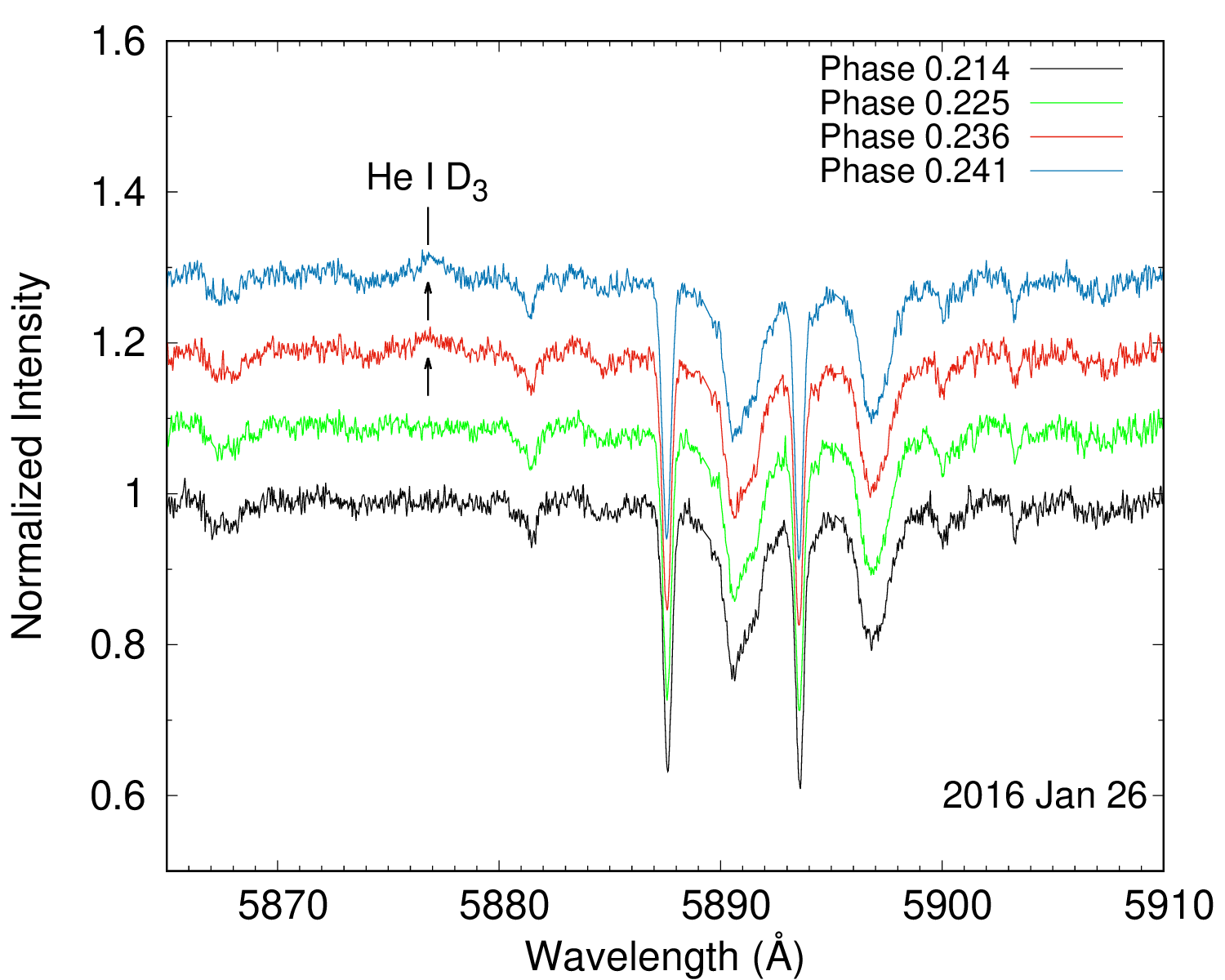}
\caption{H$_{\alpha}$ (top) and $\mbox{He~{\sc i}}$ D$_{3}$ (bottom) line profiles at the annotated phases, observed on 2016 January~26. In the top panel the subtracted spectra are shifted for better visibility. Corresponding EW variations of the subtracted H$_{\alpha}$ profiles are shown in the inset. In the bottom panel, the observed $\mbox{He~{\sc i}}$~D$_{3}$ line profiles are arranged and shifted in order of increasing phase. Arrows indicate notable $\mbox{He~{\sc i}}$~D$_{3}$ line emission features.}
\label{Fig4}
\end{figure}
\subsubsection{Prominence column density and mass}
\indent
The prominence detected on 2017 December~01 can be seen both in the H$_{\alpha}$ and H$_{\beta}$ lines, it  is much stronger than its 2017 April counterpart. Here we characterize the physical properties of this stronger prominence.\\
\indent
Assuming a prominence temperature of roughly 10,000~K, hence a thermal Doppler velocity of hydrogen of 12.0 km/s, furthermore assuming a random turbulent velocity of about 5~km/s, \citet{Dunstone2006b} used the curve of growth method to obtain the column densities of H$_{\alpha}$ in the n~=~2 level and of singly ionized $\mbox{Ca~{\sc ii}}$ atoms in the ground state. They then used the hydrogen-to-calcium ratio in solar prominences to obtain the column density of hydrogen atoms in the ground state. In this way they derived column densities of prominences on Speedy~Mic in the n~=~2 level of hydrogen with a mean value of $logN_{2}$ = 18.61~$\pm$~0.05~m$^{-2}$ and in the ground state with a mean value of $logN_{1}$ = 23.38~$\pm$~0.11~m$^{-2}$.\\
\indent
Since no $\mbox{Ca~{\sc ii}}$ H\&K measurements available, we adopt a different strategy, following \citet{Leitzinger2016}. These authors used the determined ratio of EW(H$_{\beta}$)/EW(H$_{\alpha}$) ($\sim$~0.37) in prominence absorptions to estimate the logarithmic hydrogen column density for the state $N_{2}$ from the curve of growth given in \citet{Dunstone2006b} and tentatively applied the $N_{1}$-to-$N_{2}$ ratio of Speedy~Mic to their target to derive the column density of hydrogen atoms in the ground state. They derived $logN_{2}$ and $logN_{1}$ values of about 18.0~m$^{-2}$ and 23.7~m$^{-2}$ on the fast rotating dMe star HK~Aqr, respectively.\\
 \indent 
To measure EWs of prominence absorptions in our H$_{\alpha}$ and H$_{\beta}$ profiles, we fit the subtracted spectra with multi-Gaussian profiles and then identify the Gaussian absorption component as the prominence and the Gaussian emission component as the chromospheric activity. In this way we obtained a ratio EW(H$_{\beta}$)/EW(H$_{\alpha}$) of about 0.33 and derive a $logN_{2}$ of about 17.6~m$^{-2}$ from the curve of growth of \citet{Dunstone2006b}. We then derived a $logN_{1}$ of about 22.4~m$^{-2}$ through adopting the $N_{1}$ to $N_{2}$ ratio of Speedy~Mic. These values are slightly smaller than those found by \citet{Dunstone2006b} and \citet{Leitzinger2016} for their target stars.\\
\indent
Now, having obtained the column density in the ground state of hydrogen, we can calculate the mass of prominence by using $M~=~m_{H}N_{1}A$, where $m_{H}$ is the mass of a hydrogen atom and $A$ is the projected area of the prominence. Again, we follow the method in \citet{Dunstone2006b} and \citet{Leitzinger2016} to estimate the projected prominence area. If the prominence is optically thick, when transiting the center of the stellar disk, the flux fraction of the prominence absorption can be used to calculate the fraction of the star obscured by the prominence (see the equation~(3) given in \citealt{Leitzinger2016}). For our situation, by comparing the EW(H$_{\beta}$)/EW(H$_{\alpha}$) ratio to the theoretical curve of growth of \citet{Dunstone2006b}, it can be seen that the ratio is close to the saturated part of the theoretical curve and therefore the assumption of the optically thick case seems justified. Different from Speed~Mic and HK~Aqr, RS~CVn is a binary and therefore the visible background area is the sum of two components of the system. We have calculated a mass of $2.7~\times~10^{15}$~kg for hydrogen of this prominence, which is slightly larger than the values of prominences in single active stars, which are in the range of 2--6$~\times~10^{14}$~kg for AB~Dor \citep{Collier1990}, 0.5--2.3$~\times~10^{14}$~kg for Speedy~Mic \citep{Dunstone2006b}, and $5.7~\times~10^{13}$~kg for HK~Aqr \citep{Leitzinger2016}. Our result may be reasonable, because close binary stars are usually more active than single stars and the stronger prominence absorption found in RS~CVn implies prominence material with more mass.\\
\subsection{Flare event}
\indent
Similar to the solar case, stellar flares are powerful and explosive phenomena in the outer atmosphere, commonly believed to be caused by the energy released in magnetic reconnections \citep{Schrijver2000}. Flares can be observed across the entire electromagnetic spectrum from shorter X-ray to longer radio wavelengths. Due to its very high excitation potential, the $\mbox{He~{\sc i}}$ D$_{3}$ line is an important indicator to trace flare activity in solar and stellar chromospheres. $\mbox{He~{\sc i}}$ D$_{3}$ shows obvious emission features above the continuum level during an optical flare, as widely observed on the Sun \citep{Zirin1988} and in several RS CVn-type stars like II~Peg \citep{Huenemoerder1987, Montes1997, Berdyugina1999, Frasca2008}, V711~Tau \citep{Garcia2003, Cao2015}, UX~Ari \citep{Montes1996b, Gu2002, Cao2017}, DM~UMa \citep{Zhang2016} and SZ~Psc \citep{Cao2019, Cao2020}.\\
\indent
Some consecutive H$_{\alpha}$ and $\mbox{He~{\sc i}}$ D$_{3}$ line profiles observed on 2016 January~26 are shown in Fig.~\ref{Fig4}. It can be seen that the H$_{\alpha}$ line profile shows a significant increase by a factor of about 1.3 from phase 0.225 to 0.236, and simultaneously the $\mbox{He~{\sc i}}$ D$_{3}$ line changes to emission during this phase interval. Simultaneously, other chromospheric lines also show a strengthened emission. Taking into consideration these combined facts, a flare event was detected during this night.\\
\indent
According to the measured EWs of the excess H$_{\alpha}$ emission, the observation at phase 0.236 may well correspond to the flare maximum (as also seen in Fig.~\ref{Fig4}). We compute the stellar continuum flux $F_{H_{\alpha}}$~(in erg cm$^{-2}$ s$^{-1}$ \AA$^{-1}$) near H$_{\alpha}$ as a function of the color index $B-V$ ($\sim$ 0.621 for RS~CVn; \citealt{Messina2008}) based on the empirical relationship 
\begin{eqnarray}
\log{F_{H_{\alpha}}}=[7.538-1.081(B-V)]\pm{0.33}\nonumber \\
0.0~\leq~B-V~\leq~1.4
\end{eqnarray}
of \citet{Hall1996}, and then convert the EW into an absolute surface flux $F_{S}$~(in erg~cm$^{-2}$~s$^{-1}$). Therefore, the flare emits $1.66 \times 10^{31}$ erg~s$^{-1}$ at flare maximum in H$_{\alpha}$, which is derived by converting the absolute surface flux into luminosity. During above calculation, we have corrected the EW to the total continuum before conversion to absolute flux at the stellar surface, and we assume that the flare occurred on the primary, since the intensity enhancement was associated with this star (see Fig.~\ref{Fig4}). The energy released in the H$_{\alpha}$ line is of similar order of magnitude as strong flares on other highly active RS CVn-type stars, such as V711~Tau \citep{Cao2015}, UX~Ari \citep{Montes1996b, Gu2002, Cao2017}, HK~Lac \citep{Catalano1994}, and SZ~Psc \citep{Cao2019, Cao2020}.\\
\indent
We note that, as visible in Fig.~\ref{Fig4}, during the increasing phase of the flare, the excess H$_{\alpha}$ emission profile shows a red asymmetry, i.e. the increase on the red side of the profile is stronger than on the blue, most pronounced from phase 0.225 to 0.236. This feature had also been observed during flares in the active stars PW~And \citep{Lopez-santiago2003} and LQ~Hya \citep{Montes1999}. To our knowledge, the red asymmetry is frequently seen in chromospheric activity lines during solar flares, and usually believed to result from chromospheric downward condensations \citep{Canfield1990}.\\
\begin{figure}
\centering
\includegraphics[width=8.5cm,height=6.0cm]{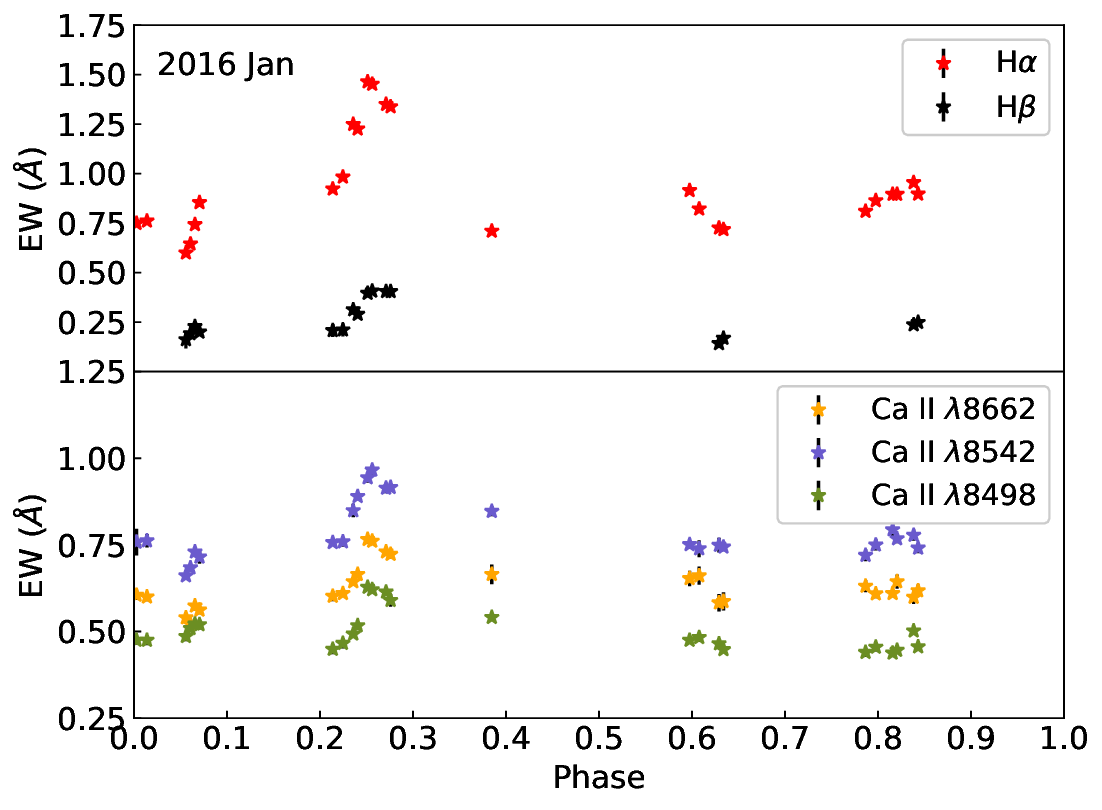}
\includegraphics[width=8.5cm,height=6.0cm]{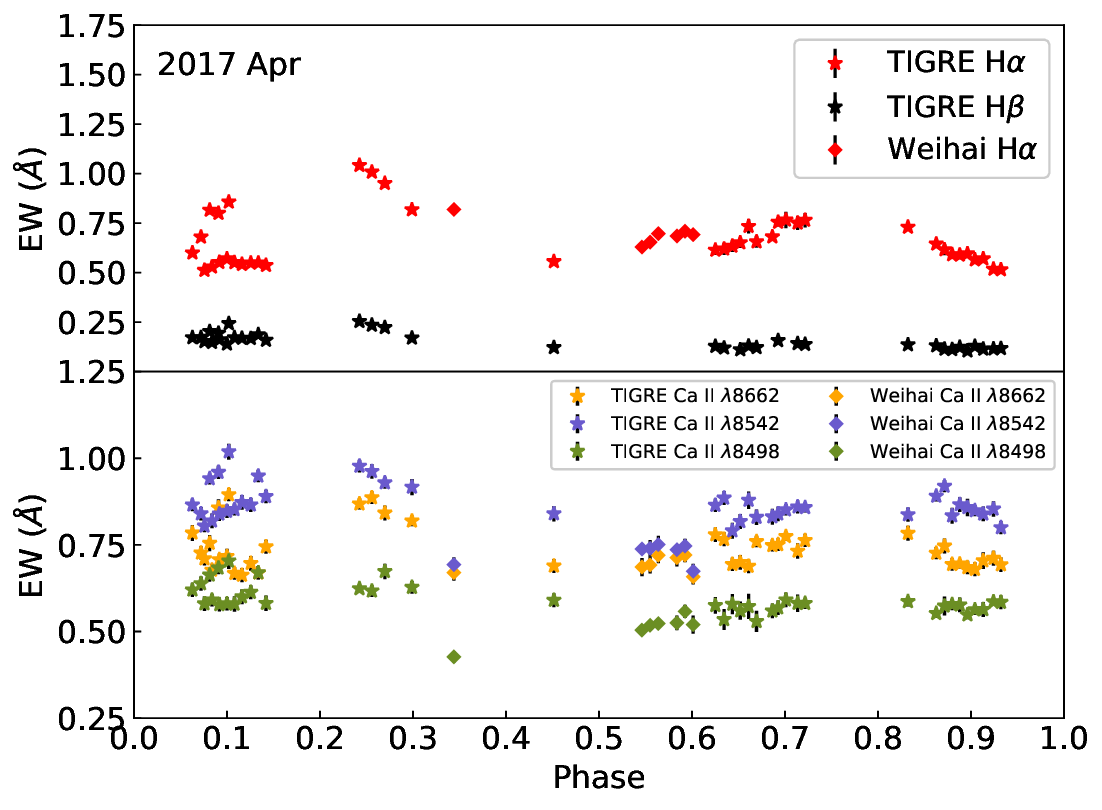}
\includegraphics[width=8.5cm,height=6.0cm]{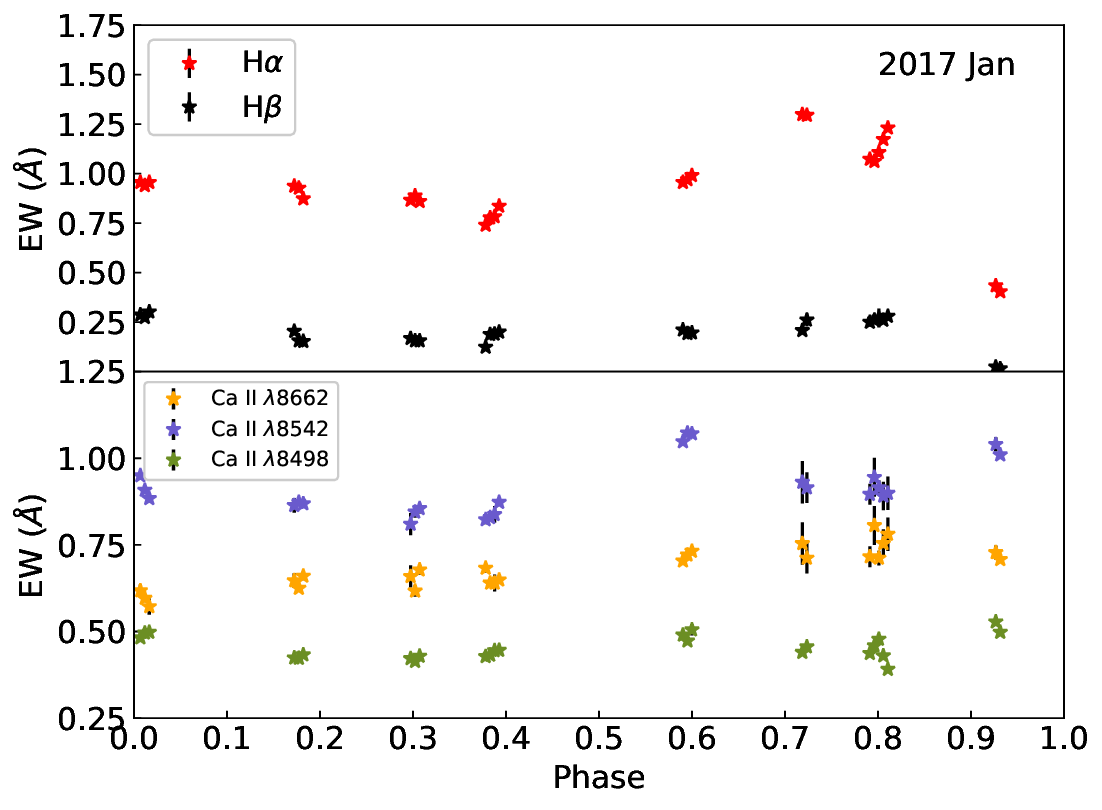}
\caption{EWs of the subtracted $\mbox{Ca~{\sc ii}}$ IRT, H$_{\alpha}$, and H$_{\beta}$ line profiles versus orbital phase, the corresponding observing runs are marked in each panel.}
\label{Fig5}
\end{figure}
\subsection{EW variations and active longitudes}
\indent
The observing runs of Jan.~2016 at Xinglong,  Apr.~2017 by the TIGRE and Weihai telescopes, as well as Nov--Dec.~2017 again at Xinglong had a denser orbital phase coverage than our earlier observations. We group those observations together to analyze a possible rotational modulation of activity, and therefore jointly analyze the EWs of H$_{\alpha}$, H$_{\beta}$, and $\mbox{Ca~{\sc ii}}$ IRT line subtractions as a function of orbital phase.\\
\indent
As shown in Fig.~\ref{Fig5}, the EW variations of the different chromospheric activity indicators are closely correlated. In 2016 January, the most striking feature is the pronounced enhancement in chromospheric emission around the first quadrature of the system, because there is a strong optical flare as discussed in Section~4.2, which possibly indicates that an active longitude exists here. For the 2017~April observing run, there are two active longitudes appearing near the two quadratures of the system where the first active longitude is stronger than the second one. For the 2017 November--December observations, the overall feature of the chromospheric variation is flat during the first half of the orbital phase,  but a relatively strong active longitude appears near the second quadrature.\\
\indent
Therefore, the chromospherically active longitudes of RS~CVn most frequently appear near the two quadratures of the system and show changes between observing runs. Similar findings were also made in other chromospheric activity stars \citep[e.g.][]{Zhang2008, Cao2020}. Based on the same datasets, moreover, \citet{Xiang2020} reconstructed surface spot maps of the primary star using Doppler imaging. The surface maps of 2016 January and 2017~April show that there are strong spot groups around phase 0.25, which are spatially associated with the chromospheric activity longitudes identified here.\\
\section{Summary and conclusions}
Based on the above analysis, our main results are:
\begin{enumerate}
\item RS~CVn shows excess emission features in the $\mbox{Ca~{\sc ii}}$ IRT, H$_{\alpha}$, and H$_{\beta}$ lines, and the chromospheric emission stems mainly from the K2~IV primary of the system, which agrees to previous results reported by other authors. In addition, the F5~V secdonary shows weak chromospheric emission in some of our observations, implying an active, albeit much less active, chromosphere on this star, too.
\item There are some unusual excess absorption features in the subtracted spectra taken near primary eclipse, which appear to be caused by two prominences located on the primary star. These tentative prominences absorb radiation from the secondary star at phases near primary eclipse. We have investigated the physical properties of the prominence structure based on the absorption features and the geometry of RS~CVn. Two prominences are estimated to have heights of 1.12~$R_{pri}$ -- 1.17~$R_{pri}$ and 0.45~$R_{pri}$ -- 1.31~$R_{pri}$ above the surface of the primary star, respectively. Moreover, we characterize the stronger to get a mass of $2.7~\times~10^{15}$~kg for hydrogen.
\item An optical flare was detected on 2016 January~26, most strongly indicated by the $\mbox{He~{\sc i}}$ D$_{3}$ line emission feature. The energy released in the H$_{\alpha}$ line during flare maximum, $1.66 \times 10^{31}$ erg~s$^{-1}$, has a similar order of magnitude as strong flares of other very active RS CVn-type stars.
\item RS~CVn shows rotational modulation of chromospheric activity, which means the presence of the chromospheric activity longitudes over the surface of the primary star. The active longitudes most frequently appear around the two quadratures of the binary system.
\end{enumerate}
\section*{Acknowledgements}
We would like to thank the staff of the telescopes for their help and support during our all observations. We are grateful to the referee Prof.~Jeffery Linsky for his valuable suggestions which result in a large improvement to our manuscript. This study make use of the data obtained with the TIGRE telescope, which located at La Luz observatory, Mexico. TIGRE is a collaboration of the Hamburger Sternwarte, the Universities of Hamburg, Guanajuato and Li\`{e}ge. This work is partially supported by the Open Project Program of the Key Laboratory of Optical Astronomy, National Astronomical Observatories, Chinese Academy of Sciences. The joint research project between Yunnan Observatories and Hamburg Observatory is funded by Sino-German Center for Research Promotion (GZ1419). The present study is also financially supported by the National Natural Science Foundation of China (NSFC) under grants Nos. 10373023, 10773027, 11333006, U1531121, and 11903074, and supported by the Natural Science Foundation of Yunnan Province of China (Grants Nos.~202201AT070186 and 202305AS350009). We acknowledge the science research grant from the China Manned Space Project with NO. CMS-CSST-2021-B07.
\section*{Data Availability}
The data underlying this article are available from the authors upon request.


\appendix
\section{EW measurements for the subtraction of chromospheric activity indicators}
\begin{table*}
\centering
\caption{EW measurements for the subtraction profiles of the $\mbox{Ca~{\sc ii}}$ $\lambda8662$, $\mbox{Ca~{\sc ii}}$ $\lambda8542$, $\lambda8498$, 
H$_{\alpha}$, and H$_{\beta}$ lines, and the ratios of EW($\lambda$8542)/EW($\lambda$8498) and E$_{H{\alpha}}$/E$_{H{\beta}}$.}
\tabcolsep 0.3cm
\label{tab3}
\begin{tabular}{@{}cccccccc}
\hline
Phase & \multicolumn{5}{c}{EW({\AA})} & $\frac{EW(\lambda8542)}{EW(\lambda8498)}$& $\frac{E_{H{\alpha}}}{E_{H{\beta}}}$ \\
\cline{2-6}\\
&$\mbox{Ca~{\sc ii}}$ $\lambda8662$ &
   $\mbox{Ca~{\sc ii}}$ $\lambda8542$ & 
   $\mbox{Ca~{\sc ii}}$ $\lambda8498$ & 
   H$_{\alpha}$ & 
   H$_{\beta}$ & \\
\hline
 \multicolumn{8}{c}{1998 Mar, Xinglong~2.16~m}\\
0.776&0.697$\pm$0.045&0.810$\pm$0.015&...&0.692$\pm$0.019&...&...&...\\
0.785&0.664$\pm$0.017&0.787$\pm$0.013&...&0.709$\pm$0.011&...&...&...\\
\hline
 \multicolumn{8}{c}{2000 Feb, Xinglong~2.16~m}\\
0.346&0.539$\pm$0.015&0.736$\pm$0.015&0.476$\pm$0.012&0.842$\pm$0.015&...&1.55&...\\
0.350&0.560$\pm$0.012&0.788$\pm$0.019&0.509$\pm$0.005&0.867$\pm$0.013&...&1.55&...\\
\hline
 \multicolumn{8}{c}{2004 Feb, Xinglong~2.16~m}\\
0.341&0.567$\pm$0.018&0.782$\pm$0.016&0.428$\pm$0.009&0.648$\pm$0.012&...&1.83&...\\
0.734&0.518$\pm$0.006&0.833$\pm$0.017&0.559$\pm$0.018&0.478$\pm$0.015&...&1.49&...\\
0.752&0.492$\pm$0.016&0.758$\pm$0.011&0.500$\pm$0.009&0.563$\pm$0.013&...&1.52&...\\
0.160&0.516$\pm$0.011&0.767$\pm$0.021&0.510$\pm$0.008&0.932$\pm$0.019&...&1.50&...\\
0.380&0.608$\pm$0.012&0.856$\pm$0.015&0.483$\pm$0.008&0.685$\pm$0.011&...&1.77&...\\
0.576&0.557$\pm$0.012&0.689$\pm$0.020&0.516$\pm$0.011&0.452$\pm$0.021&...&1.34&...\\
\hline
 \multicolumn{8}{c}{2016 Jan, Xinglong~2.16~m}\\
0.385&0.597$\pm$0.018&0.847$\pm$0.014&0.541$\pm$0.009&0.710$\pm$0.018&...&1.57&...\\
0.598&0.669$\pm$0.013&0.751$\pm$0.012&0.475$\pm$0.008&0.915$\pm$0.010&...&1.58&...\\
0.608&0.682$\pm$0.008&0.739$\pm$0.024&0.483$\pm$0.013&0.822$\pm$0.011&...&1.53&...\\
0.787&0.631$\pm$0.017&0.720$\pm$0.017&0.440$\pm$0.013&0.810$\pm$0.011&...&1.64&...\\
0.798&0.609$\pm$0.013&0.750$\pm$0.016&0.455$\pm$0.012&0.864$\pm$0.014&...&1.65&...\\
0.816&0.610$\pm$0.011&0.794$\pm$0.018&0.438$\pm$0.011&0.896$\pm$0.006&...&1.81&...\\
0.821&0.644$\pm$0.020&0.768$\pm$0.013&0.446$\pm$0.006&0.896$\pm$0.010&...&1.72&...\\
0.214&0.602$\pm$0.014&0.757$\pm$0.010&0.449$\pm$0.014&0.923$\pm$0.014&0.208$\pm$0.032&1.69&2.29\\
0.225&0.610$\pm$0.009&0.759$\pm$0.014&0.466$\pm$0.015&0.983$\pm$0.009&0.211$\pm$0.033&1.63&2.40\\
0.236&0.644$\pm$0.012&0.849$\pm$0.019&0.493$\pm$0.009&1.249$\pm$0.012&0.313$\pm$0.004&1.72&2.06\\
0.241&0.665$\pm$0.010&0.890$\pm$0.011&0.517$\pm$0.008&1.226$\pm$0.010&0.290$\pm$0.008&1.72&2.18\\
0.629&0.564$\pm$0.013&0.749$\pm$0.021&0.465$\pm$0.009&0.726$\pm$0.009&0.142$\pm$0.010&1.61&2.63\\
0.634&0.566$\pm$0.009&0.744$\pm$0.011&0.448$\pm$0.010&0.718$\pm$0.008&0.169$\pm$0.012&1.66&2.19\\
0.838&0.599$\pm$0.019&0.778$\pm$0.016&0.502$\pm$0.012&0.955$\pm$0.009&0.237$\pm$0.005&1.55&2.08\\
0.843&0.618$\pm$0.017&0.741$\pm$0.015&0.456$\pm$0.009&0.897$\pm$0.015&0.249$\pm$0.015&1.63&1.58\\
0.056&0.540$\pm$0.012&0.661$\pm$0.012&0.486$\pm$0.014&0.600$\pm$0.014&0.162$\pm$0.044&1.36&1.91\\
0.061&0.521$\pm$0.006&0.684$\pm$0.011&0.509$\pm$0.010&0.645$\pm$0.006&0.192$\pm$0.006&1.34&1.73\\
0.066&0.574$\pm$0.013&0.730$\pm$0.013&0.522$\pm$0.006&0.743$\pm$0.011&0.228$\pm$0.012&1.40&1.68\\
0.071&0.562$\pm$0.012&0.715$\pm$0.019&0.520$\pm$0.014&0.854$\pm$0.012&0.200$\pm$0.010&1.38&2.20\\
0.251&0.766$\pm$0.012&0.944$\pm$0.014&0.628$\pm$0.011&1.464$\pm$0.013&0.396$\pm$0.006&1.50&1.91\\
0.256&0.761$\pm$0.007&0.966$\pm$0.013&0.621$\pm$0.011&1.452$\pm$0.013&0.407$\pm$0.012&1.56&1.84\\
0.271&0.730$\pm$0.010&0.914$\pm$0.015&0.614$\pm$0.010&1.350$\pm$0.008&0.405$\pm$0.007&1.49&1.72\\
0.276&0.723$\pm$0.012&0.916$\pm$0.011&0.590$\pm$0.018&1.338$\pm$0.009&0.405$\pm$0.013&1.55&1.70\\
\hline
\multicolumn{8}{c}{2017 Apr, Weihai~1~m}\\
0.344&0.670$\pm$0.023&0.693$\pm$0.021&0.427$\pm$0.016&0.819$\pm$0.011&...&1.62&...\\
0.546&0.686$\pm$0.026&0.738$\pm$0.015&0.504$\pm$0.012&0.629$\pm$0.015&...&1.46&...\\
0.555&0.692$\pm$0.024&0.740$\pm$0.024&0.518$\pm$0.014&0.654$\pm$0.007&...&1.43&...\\
0.564&0.720$\pm$0.023&0.751$\pm$0.025&0.523$\pm$0.015&0.697$\pm$0.015&...&1.44&...\\
0.584&0.713$\pm$0.025&0.736$\pm$0.015&0.525$\pm$0.021&0.685$\pm$0.015&...&1.40&...\\
0.593&0.721$\pm$0.021&0.747$\pm$0.020&0.558$\pm$0.019&0.709$\pm$0.026&...&1.34&...\\
0.601&0.658$\pm$0.022&0.674$\pm$0.021&0.520$\pm$0.026&0.692$\pm$0.012&...&1.30&...\\
\hline
\multicolumn{8}{c}{2017 Apr, TIGRE~1.2~m}\\
0.625&0.780$\pm$0.021&0.864$\pm$0.019&0.575$\pm$0.024&0.614$\pm$0.024&0.128$\pm$0.015&1.50&2.47\\
0.635&0.766$\pm$0.020&0.885$\pm$0.020&0.535$\pm$0.030&0.620$\pm$0.030&0.120$\pm$0.017&1.65&2.66\\
0.644&0.694$\pm$0.019&0.791$\pm$0.023&0.579$\pm$0.029&0.634$\pm$0.029&...&1.37&...\\
0.652&0.699$\pm$0.021&0.818$\pm$0.021&0.560$\pm$0.026&0.651$\pm$0.026&0.111$\pm$0.013&1.46&3.02\\
0.661&0.689$\pm$0.020&0.879$\pm$0.025&0.572$\pm$0.036&0.734$\pm$0.036&0.130$\pm$0.011&1.54&2.91\\
0.670&0.760$\pm$0.016&0.830$\pm$0.022&0.529$\pm$0.030&0.656$\pm$0.030&0.122$\pm$0.017&1.57&2.77\\
0.687&0.749$\pm$0.017&0.831$\pm$0.023&0.560$\pm$0.021&0.681$\pm$0.021&...&1.48&...\\
0.832&0.784$\pm$0.021&0.837$\pm$0.020&0.587$\pm$0.014&0.730$\pm$0.014&0.136$\pm$0.015&1.43&2.77\\
0.063&0.784$\pm$0.023&0.865$\pm$0.018&0.620$\pm$0.020&0.600$\pm$0.012&0.172$\pm$0.013&1.40&1.80\\
0.072&0.727$\pm$0.019&0.840$\pm$0.019&0.638$\pm$0.020&0.680$\pm$0.008&0.172$\pm$0.010&1.32&2.04\\
0.081&0.755$\pm$0.023&0.941$\pm$0.017&0.663$\pm$0.019&0.816$\pm$0.020&0.203$\pm$0.009&1.42&2.07\\
\end{tabular}
\end{table*}
\begin{table*}
\centering
\contcaption{}
\tabcolsep 0.3cm
\label{tab3}
\begin{tabular}{@{}cccccccc}
\hline
Phase & \multicolumn{5}{c}{EW({\AA})} & $\frac{EW(\lambda8542)}{EW(\lambda8498)}$& $\frac{E_{H{\alpha}}}{E_{H{\beta}}}$ \\
\cline{2-6}\\
&$\mbox{Ca~{\sc ii}}$ $\lambda8662$ &
   $\mbox{Ca~{\sc ii}}$ $\lambda8542$ & 
   $\mbox{Ca~{\sc ii}}$ $\lambda8498$ & 
   H$_{\alpha}$ & 
   H$_{\beta}$ & \\
\hline
0.091&0.857$\pm$0.024&0.960$\pm$0.019&0.684$\pm$0.018&0.800$\pm$0.012&0.194$\pm$0.011&1.40&2.13\\
0.102&0.895$\pm$0.019&1.019$\pm$0.023&0.704$\pm$0.023&0.857$\pm$0.013&0.243$\pm$0.013&1.45&1.82\\
0.243&0.868$\pm$0.016&0.978$\pm$0.018&0.624$\pm$0.012&1.042$\pm$0.012&0.254$\pm$0.012&1.57&2.11\\
0.256&0.886$\pm$0.018&0.962$\pm$0.021&0.617$\pm$0.017&1.008$\pm$0.017&0.235$\pm$0.011&1.56&2.21\\
0.270&0.842$\pm$0.021&0.930$\pm$0.017&0.674$\pm$0.022&0.951$\pm$0.022&0.223$\pm$0.012&1.38&2.20\\
0.299&0.819$\pm$0.015&0.917$\pm$0.023&0.628$\pm$0.018&0.818$\pm$0.018&0.170$\pm$0.015&1.46&2.48\\
0.452&0.689$\pm$0.020&0.840$\pm$0.022&0.590$\pm$0.020&0.556$\pm$0.011&0.122$\pm$0.011&1.42&2.35\\
0.693&0.751$\pm$0.018&0.840$\pm$0.021&0.569$\pm$0.021&0.755$\pm$0.021&0.157$\pm$0.013&1.48&2.48\\
0.701&0.775$\pm$0.013&0.851$\pm$0.016&0.591$\pm$0.021&0.766$\pm$0.041&...&1.44&...\\
0.714&0.732$\pm$0.022&0.860$\pm$0.019&0.582$\pm$0.025&0.751$\pm$0.035&0.142$\pm$0.012&1.48&2.73\\
0.722&0.763$\pm$0.019&0.858$\pm$0.014&0.581$\pm$0.016&0.765$\pm$0.036&0.138$\pm$0.017&1.48&2.86\\
0.863&0.726$\pm$0.017&0.891$\pm$0.015&0.552$\pm$0.013&0.645$\pm$0.013&0.131$\pm$0.013&1.61&2.54\\
0.872&0.747$\pm$0.021&0.920$\pm$0.017&0.574$\pm$0.027&0.616$\pm$0.027&0.114$\pm$0.015&1.60&2.78\\
0.880&0.694$\pm$0.018&0.834$\pm$0.023&0.578$\pm$0.016&0.591$\pm$0.016&0.112$\pm$0.011&1.44&2.72\\
0.888&0.695$\pm$0.013&0.866$\pm$0.021&0.577$\pm$0.021&0.589$\pm$0.021&0.123$\pm$0.013&1.50&2.47\\
0.896&0.685$\pm$0.019&0.858$\pm$0.025&0.548$\pm$0.013&0.596$\pm$0.013&0.105$\pm$0.010&1.57&2.93\\
0.904&0.680$\pm$0.020&0.849$\pm$0.016&0.563$\pm$0.015&0.565$\pm$0.015&0.128$\pm$0.011&1.51&2.27\\
0.913&0.704$\pm$0.024&0.839$\pm$0.021&0.564$\pm$0.021&0.570$\pm$0.021&0.114$\pm$0.013&1.49&2.58\\
0.924&0.711$\pm$0.021&0.854$\pm$0.022&0.586$\pm$0.011&0.518$\pm$0.011&0.115$\pm$0.017&1.46&2.32\\
0.932&0.693$\pm$0.021&0.800$\pm$0.019&0.584$\pm$0.020&0.515$\pm$0.013&0.117$\pm$0.013&1.37&2.27\\
0.076&0.710$\pm$0.019&0.805$\pm$0.018&0.580$\pm$0.021&0.512$\pm$0.010&0.152$\pm$0.012&1.39&1.74\\
0.084&0.677$\pm$0.018&0.819$\pm$0.021&0.591$\pm$0.018&0.529$\pm$0.011&0.148$\pm$0.011&1.39&1.84\\
0.092&0.708$\pm$0.017&0.839$\pm$0.020&0.578$\pm$0.020&0.552$\pm$0.009&0.162$\pm$0.015&1.45&1.76\\
0.100&0.717$\pm$0.022&0.847$\pm$0.019&0.580$\pm$0.019&0.570$\pm$0.010&0.140$\pm$0.010&1.46&2.10\\
0.108&0.668$\pm$0.018&0.852$\pm$0.017&0.579$\pm$0.022&0.551$\pm$0.021&0.172$\pm$0.012&1.47&1.65\\
0.116&0.663$\pm$0.020&0.873$\pm$0.021&0.600$\pm$0.021&0.542$\pm$0.016&0.169$\pm$0.011&1.46&1.65\\
0.126&0.697$\pm$0.021&0.866$\pm$0.020&0.613$\pm$0.020&0.549$\pm$0.016&0.168$\pm$0.013&1.41&1.68\\
0.134&0.669$\pm$0.019&0.949$\pm$0.018&0.670$\pm$0.019&0.549$\pm$0.011&0.187$\pm$0.016&1.42&1.51\\
0.142&0.745$\pm$0.020&0.890$\pm$0.019&0.581$\pm$0.022&0.537$\pm$0.010&0.159$\pm$0.012&1.53&1.74\\
\hline
\multicolumn{8}{c}{2017 Nov--Dec, Xinglong~2.16~m}\\
0.298&0.659$\pm$0.032&0.810$\pm$0.032&0.422$\pm$0.014&0.866$\pm$0.007&0.168$\pm$0.012&1.92&2.65\\
0.302&0.617$\pm$0.016&0.845$\pm$0.016&0.414$\pm$0.018&0.888$\pm$0.019&0.156$\pm$0.015&2.04&2.93\\
0.307&0.678$\pm$0.014&0.855$\pm$0.014&0.429$\pm$0.016&0.860$\pm$0.017&0.156$\pm$0.015&1.99&2.84\\
0.719&0.754$\pm$0.061&0.931$\pm$0.061&0.440$\pm$0.015&1.299$\pm$0.012&0.208$\pm$0.011&2.12&3.22\\
0.724&0.712$\pm$0.044&0.915$\pm$0.044&0.456$\pm$0.014&1.295$\pm$0.012&0.261$\pm$0.032&2.01&2.56\\
0.927&0.728$\pm$0.020&1.040$\pm$0.020&0.528$\pm$0.015&0.434$\pm$0.015&0.024$\pm$0.008&1.97&9.32\\
0.932&0.708$\pm$0.008&1.010$\pm$0.008&0.498$\pm$0.016&0.404$\pm$0.010&0.015$\pm$0.004&2.03&13.88\\
0.172&0.647$\pm$0.021&0.864$\pm$0.021&0.424$\pm$0.013&0.937$\pm$0.009&0.204$\pm$0.011&2.04&2.37\\
0.177&0.625$\pm$0.006&0.873$\pm$0.006&0.423$\pm$0.014&0.927$\pm$0.012&0.156$\pm$0.022&2.06&3.06\\
0.182&0.660$\pm$0.012&0.869$\pm$0.012&0.433$\pm$0.014&0.873$\pm$0.008&0.153$\pm$0.018&2.01&2.94\\
0.378&0.683$\pm$0.013&0.823$\pm$0.013&0.428$\pm$0.014&0.740$\pm$0.020&0.124$\pm$0.011&1.92&3.08\\
0.383&0.640$\pm$0.002&0.829$\pm$0.002&0.432$\pm$0.012&0.777$\pm$0.012&0.188$\pm$0.010&1.92&2.13\\
0.388&0.640$\pm$0.024&0.838$\pm$0.024&0.445$\pm$0.018&0.783$\pm$0.013&0.190$\pm$0.013&1.88&2.12\\
0.393&0.649$\pm$0.007&0.873$\pm$0.007&0.446$\pm$0.013&0.836$\pm$0.005&0.200$\pm$0.025&1.96&2.15\\
0.590&0.704$\pm$0.008&1.048$\pm$0.008&0.490$\pm$0.018&0.956$\pm$0.008&0.211$\pm$0.010&2.14&2.33\\
0.595&0.721$\pm$0.008&1.073$\pm$0.008&0.473$\pm$0.014&0.970$\pm$0.012&0.193$\pm$0.008&2.27&2.59\\
0.600&0.732$\pm$0.008&1.071$\pm$0.008&0.505$\pm$0.016&0.991$\pm$0.009&0.196$\pm$0.007&2.12&2.61\\
0.791&0.716$\pm$0.030&0.896$\pm$0.030&0.437$\pm$0.012&1.073$\pm$0.009&0.250$\pm$0.023&2.05&2.21\\
0.796&0.806$\pm$0.056&0.945$\pm$0.056&0.459$\pm$0.014&1.062$\pm$0.011&0.261$\pm$0.020&2.06&2.10\\
0.801&0.712$\pm$0.021&0.915$\pm$0.021&0.478$\pm$0.013&1.107$\pm$0.016&0.271$\pm$0.047&1.91&2.10\\
0.806&0.754$\pm$0.041&0.891$\pm$0.041&0.430$\pm$0.015&1.173$\pm$0.012&0.260$\pm$0.011&2.07&2.32\\
0.81&0.781$\pm$0.048&0.899$\pm$0.048&0.391$\pm$0.017&1.231$\pm$0.007&0.280$\pm$0.013&2.30&2.27\\
0.007&1.085$\pm$0.017&0.950$\pm$0.008&0.482$\pm$0.013&0.956$\pm$0.018&0.287$\pm$0.021&1.97&1.72\\
0.012&1.045$\pm$0.011&0.908$\pm$0.009&0.496$\pm$0.012&0.940$\pm$0.008&0.274$\pm$0.013&1.83&1.77\\
0.017&1.003$\pm$0.023&0.884$\pm$0.011&0.498$\pm$0.012&0.956$\pm$0.021&0.302$\pm$0.017&1.78&1.63\\
\hline
\end{tabular}
\end{table*}
\bsp	
\label{lastpage}
\end{document}